# Large Language Models, and LLM-Based Agents, Should Be Used to Enhance the Digital Public Sphere

Seth Lazar   Luke Thorburn   Tian Jin   Luca Belli
ANU          KCL             MIT        Sator Labs

**Abstract**

This paper argues that large language model-based recommenders can displace today's attention-allocation machinery. LLM-based recommenders would ingest open-web content, infer a user's natural-language goals, and present information that matches their reflective preferences. Properly designed, they could deliver personalization without industrial-scale data hoarding, return control to individuals, optimize for genuine ends rather than click-through proxies, and support autonomous attention management. Synthesizing evidence of current systems' harms with recent work on LLM-driven pipelines, we identify four key research hurdles: generating candidates without centralized data, maintaining computational efficiency, modeling preferences robustly, and defending against prompt-injection. None looks prohibitive; surmounting them would steer the digital public sphere toward democratic, human-centered values.

# 1   Introduction

The shortcomings of contemporary online communication are well documented [15, 38]. As social media ecosystems splinter, indicators of abuse, epistemic pollution, manipulation and polarization continue to rise [91]. Consequently, the digital public sphere falls well below the optimistic projections of the early 'wealth of networks' thesis [3, 15, 55].

The causes of these problems are contested. Some scholars attribute them to human communicative behavior [90, 1]; others implicate the technical architectures that mediate online interaction. A decisive answer is improbable. Yet, while large-scale behavioral change is elusive, technical systems can be redesigned.

In this paper, we therefore concentrate on the technologies that allocate online attention. Regardless of whether they are the root cause of current pathologies, these systems are entangled with practices that are independently objectionable. This observation motivates a search for alternatives.

**We contend that prevailing recommender pipelines rely on mass surveillance, reinforce concentrated platform power, embody narrow behaviorist**

assumptions, and erode user agency. **We advocate an alternative paradigm that leverages advances in large language models (LLMs) and LLM-based agents.**[1] **By representing both content and user values in natural language and invoking external tools when needed, such systems could avoid these four pathologies.**

Avoiding these harms will not guarantee a healthier digital public sphere. If underlying social dynamics drive toxicity, improved recommendation may have limited effect, and commercial incentives could push practitioners to graft LLM recommenders onto existing surveillance-based models [78, 47, 48, 79]. Nevertheless, LLM-based recommenders can be deployed outside incumbent platforms, offering researchers and developers a plausible route around current gatekeepers. We therefore propose that technical researchers motivated to improve the digital public sphere should direct their attention to this approach.

Section 2 surveys the current state of online communication. Section 3 analyses how mainstream recommenders enable the four harms above. Section 4 reviews incremental responses and introduces LLM-based recommendation. Section 5 explains how LLM systems might mitigate the identified problems. Section 6 outlines outstanding (but surmountable) technical and ethical challenges; Section 7 concludes.

## 2 Diagnosing the Digital Public Sphere

The digital public sphere—the social, cultural, and political exchanges hosted online—can be evaluated along two axes. First, by the harms it enables; second, by the positive ideals it fails to achieve.

**Harms**. Following Lazar [55], these harms cluster into three categories

- *Abuse*: direct and indirect harassment, silencing, and other forms of targeted hostility [107, 86].

- *Epistemic pollution*: practices that obstruct accurate belief-formation—mis- and disinformation, coordinated inauthentic behavior, 'flood-the-zone' tactics, SEO-driven content farms, generative-AI 'slop,' and the 'liar's dividend' [36, 106, 46, 42].

- *Manipulation*: communicative strategies that compromise agency, ranging from individual grooming and radicalization to population-level shifts such as affective polarization or body-image anxiety [93, 4, 84, 16].

**Positive ideals**. A healthy public sphere is not merely the absence of pathology; it realizes communicative goods—equitably satisfying people's interests in sharing, receiving, and deliberating information. We label this objective *communicative justice* [55]. Designing for justice therefore requires technologies that advance positive aims as well as mitigate harm.

Attention in the digital public sphere is currently allocated by deep reinforcement learning-based recommender systems [67, 96, 68]. On large platforms (e.g., Facebook, Instagram, YouTube, X, TikTok), recommender pipelines typically:

- Retrieve a candidate set via lightweight heuristics.

- Rank candidates with a computationally heavier value model—usually a weighted sum of predicted engagement metrics, survey-based quality signals, and policy classifier scores.

- Re-rank to enforce secondary constraints (e.g., diversity across consecutive items, or fairness to producers, as in Spotify or YouTube—[65, 66, 88, 100, 9, 6]).

---

[1] Note that we are using 'LLM' here to mean any transformer-based token-sequence predictor, including multimodal foundation models, vision-language models, and others.

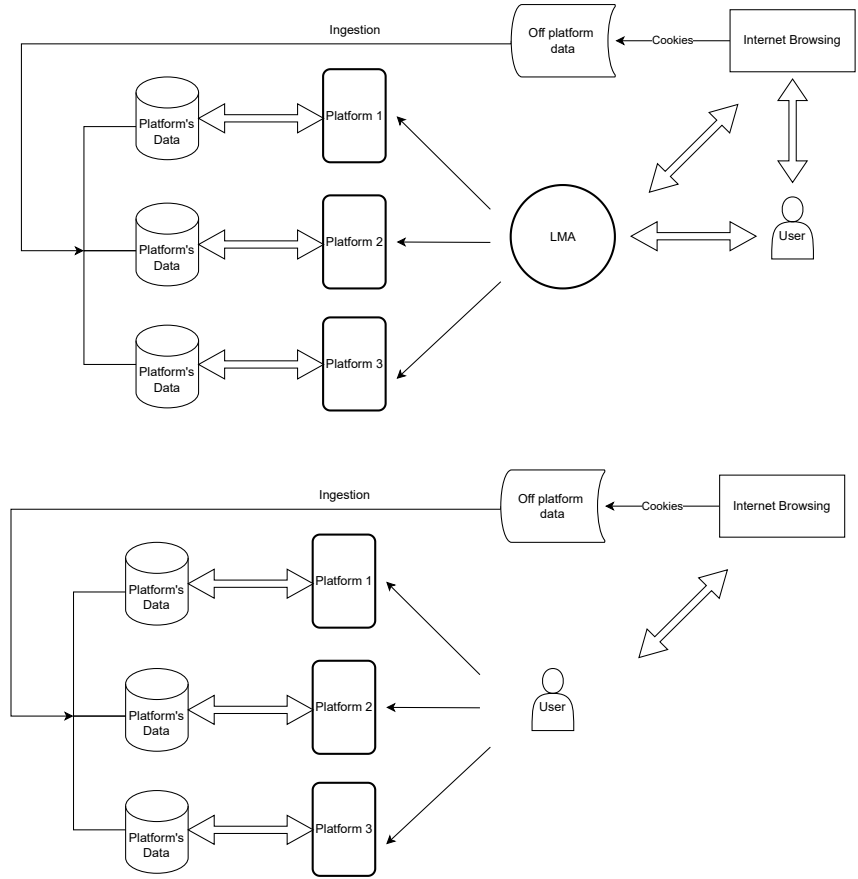

Figure 1: A schematic showing the core idea of using a Language Model Agent (LMA) as a recommender (top), compared to the current paradigm (below). In contrast to the current paradigm, an LMA could interact with web content (including but not only social media content) to source candidate items to recommend.

Because recommenders mediate much of the digital public sphere, they are attractive scapegoats for its shortcomings [25, 80]. Yet rigorous evidence for their causal role remains mixed [70, 33, 37, 72]. Decades of scholarship in science and technology studies (STS) also caution against technological determinism—the thesis that a technology alone can strictly determine significant societal outcomes. In general, societal impacts derive from complex interactions among technology, users, and social context [20].

## 2.1 A Narrow Target

Our aim is therefore limited. We identify four structural failings that today's recommender paradigm tends to reinforce—**mass surveillance**, **power concentration**, **narrow behaviorism**, and **diminished user agency**—and argue that recommender systems built around LMs, especially agentic variants, could be designed to perform the same function while avoiding these specific pathologies. Crucially, such systems need not depend on the cooperation of incumbent gatekeepers. Even if their deployment leaves other problems of the digital public sphere untouched, reducing surveillance, countering concentration, broadening behavioral assumptions, and restoring user agency are worth attempting for their own sake.

# 3 Four Problems with Recommender Systems

## 3.1 Dependence on Mass Surveillance

At a functional level, any recommender must (1) represent the items competing for a user's attention, (2) model the user's preferences and values, and (3) select and order items so as to serve those preferences. The dominant industrial approach satisfies all three requirements by observing user behavior at scale:

- **Content understanding**. Recommenders infer an item's salient properties from how large numbers of users interact with it.

- **User modeling**. They characterize a given user by comparing that user's past engagement to aggregated behavioral profiles of others.

- **Inference and ranking**. They combine both behavioral streams to predict which items will maximize a proprietary 'value function'—typically a weighted sum of engagement metrics, predicted surveys, and policy classifier scores.

Without continuous, fine-grained logging of impressions, clicks, likes, and dwell time, these systems degrade sharply. This was true for early click-through-rate pipelines that shaped Facebook's ad placement and News Feed [34, 39]. It remains true for modern value-model architectures, regardless of whether they use supervised, unsupervised, or reinforcement learning.

A large literature treats pervasive tracking as presumptively objectionable [114, 4, 103]. If current recommenders require such tracking, we should explore designs that do not. It's true that techniques like differential privacy, federated learning, and on-device inference can mask individual records. They do not remedy the collective and structural harms of mass surveillance—such as concentration of informational power or the chilling effects of ubiquitous monitoring [95, 4, 97, 87].

## 3.2 Incentivizing Power Concentration

Existing recommender systems infer content representations, user preferences, and item-to-user match quality from behavioral data. The more such data they observe, the more accurate their models become. Crucially, obtaining that data at scale is feasible only when content distribution is centralized: platform owners must intermediate most user interactions so that every click, dwell, and share flows back into their training logs.

This feedback loop creates a clear strategic objective: maximize platform scale. More users and more content yield more data; more data improves recommendation quality; higher quality attracts still more users. Scale translates directly into concentrated power: the larger the platform, the more data is being collected, the more users' attention is being allocated, the greater the degree of influence over people's welfare, options, and attitudes [58] that sits with those who own the platform and its recommenders. The result: a small set of firms control the attention of billions, determining which signals are amplified and which are suppressed [32].

Philosophical critiques of such concentration of power over attention are well rehearsed [53, 94, 55], but recent events make the practical risks equally vivid. Entrusting the global flow of information to a handful of proprietors—sometimes idiosyncratic billionaires—invites arbitrary interventions in public discourse.

One response, middleware, would let third-party recommenders sit between the platform and the user [29]. In principle, a market of competing ranking services could break the present oligopoly over attention. In practice, middleware adoption has largely stalled (notwithstanding the existence of 'custom feeds' on Bluesky) because behavioral data remain siloed. Platforms cite both competitive advantage and data-protection law when refusing to share those logs [52].

Hence, addressing power concentration requires recommender paradigms that do not depend on privileged access to platform-scale behavioral data. We turn to that possibility next.

## 3.3 Narrow Behaviorism

Current recommender systems infer user interests primarily from observable behavior—mouse-overs, scroll pauses, clicks, likes, reposts, and related signals [66, 100]. Although some contextual cues (e.g., platform assessments of content quality) are sometimes incorporated [18], behavioral data remain dominant. This emphasis produces narrowly behaviorist inferences that create at least two problems.

First, observed behavior tracks revealed rather than considered preferences [67]. Revealed preferences are the choices individuals make within the options they notice; considered preferences are the choices they would make with adequate information, a richer option set, and sufficient self-control. Because these conditions are often unmet [83], behavior is an unreliable proxy for considered preferences. For example, many users lament the heightened moral temperature of online discourse [84] yet still engage with outrage-inducing content [7, 8]. In the moment they may underestimate its cumulative costs, lack viable alternatives, or simply succumb to immediate impulses.

Second, a strictly behaviorist approach leaves scant room for encoding societal values [5]. These societal values extend beyond any one user's short-term incentives and rarely manifest as discrete choices that a recommender can observe. We may, for instance, endorse a public sphere that supports democratic deliberation, yet be offered no individual action whose click-through meaningfully signals that commitment. Optimizing for short-run individual engagement therefore tends to amplify self-interested behavior and reproduce collective-action failures that most users would reject ex ante.

## 3.4 Compromised Attentional Agency

Current recommender systems are optimized to satisfy users' immediate attentional impulses with minimal explicit input. They observe behavioral signals—pauses, mouse-overs, clicks, likes, reposts—and, by leveraging fast, automatic ('System I') cognitive processes [49], deliver a stream of content that can be consumed with little deliberation. Because these models learn almost exclusively from such passively collected traces, users cannot easily fine-tune them; meaningful 'training' requires extended usage [27] or coarse interventions such as muting keywords. Even where interface controls exist, adoption is typically limited to a small minority of users [48, 18]. The underlying high-dimensional representations are difficult to interpret, and external correction is currently infeasible [56].

These properties constrain *attentional agency*: the deliberate, reasons-responsive allocation of attention [105, 82]. While users *can* switch platforms or employ alternative discovery mechanisms, heavy reliance on opaque, hard-to-steer recommenders makes such agency more costly and so less frequent. Our claim is not that such systems eliminate agency altogether (see e.g. Hari [40]—but that they predictably weaken it.

Perceptions of weakened agency matter as well. When observers believe their interlocutors' attention is mechanically directed by algorithms, they may attribute disagreement to manipulation rather than sincere difference of opinion, thereby fueling distrust [4].

In sum, contemporary recommenders incentivize large-scale behavioral surveillance, concentrate informational power, encode narrow behaviorism, and erode attentional agency. They are not the sole cause of dysfunction in the digital public sphere, but their design choices incentivize specific, avoidable harms. Developing alternative approaches that achieve comparable utility without these costs is therefore a worthwhile research goal.

## 4  What Paths Forward?

Existing work suggests at least three responses to the shortcomings of algorithmic recommendation [110]. One is to abandon ML-based recommendation entirely. Another keeps the basic architecture of existing approaches, but aims to identify better behavioral signals. And we will introduce the third at the end of this section.

### 4.1  'Algorithms Ruin Everything' [24]

A prominent reform proposal—championed on 'fediverse' platforms such as Mastodon and Bluesky—eschews algorithmic recommendation in favor of a purely subscription- and social-network-driven model [10, 71]. Users encounter a post only if they have explicitly subscribed to its source or if it is reposted by an account they follow, with exposure mediated through their local social graph [26]. Items are then displayed in strict reverse-chronological order: an ordering rule that is simple and transparent, and does not predict relevance.

Although this design avoids certain pathologies of predictive ranking, dispensing with recommender algorithms altogether imposes substantial costs. First, it shifts the cognitive burden of discovery and prioritization onto each individual user—they demand *too much* attentional agency. Without automated ranking, feeds can become unmanageable as network size grows, and the only available controls are coarse, reactive measures (e.g., muting reposts or maintaining multiple curated lists). Second, the absence of relevance ordering can hinder real-time dialogue, especially across time-zone boundaries, because recent posts crowd out contextually salient—but slightly older—content.

In short, a feed determined solely by subscriptions preserves user autonomy but provides limited support for efficient discovery and attention management, especially the allocation of *collective* attention. A wholesale rejection of algorithmic recommendation therefore risks undermining the very communicative goals that social platforms are meant to serve.

### 4.2  Complementing Behaviorism

A more incremental strategy seeks to steer current recommender systems toward socially beneficial objectives by refining the metrics they optimize or the signals they incorporate [18]. Two avenues illustrate this approach:

- Richer explicit feedback. Expanding the range of user reactions—Facebook's shift from a single 'like' to multiple emotional responses is a canonical example—supplies the algorithm with finer-grained preferences, giving users greater influence over their feeds.

- Proxy signals for civic value. Platforms can elevate content that performs well on metrics correlated with epistemic or social goods. X's Community Notes, for instance, ranks annotations partly by whether they win approval from ideologically diverse reviewers, operationalizing a 'bridging' heuristic [75].

We do not attempt a comprehensive survey here, noting only two considerations. First, many such proposals merit serious experimentation. Second, as long as recommender systems remain grounded in large-scale behavioral surveillance and deep-learning inference controlled by centralized platforms, they will continue to incentivize data extraction, concentrate power, exhibit elements of narrow behaviorism, and constrain user agency. Mitigating these deeper structural concerns may therefore require more radical departures from the current paradigm.

## 4.3 Language Models

Large language models have been used in recommendation tasks for a while as support for the heavy ranking tasks, mainly in the form of contextual embeddings for users' or items' text features. In recent times, however, their increased capabilities have opened new research directions, including the option to use such systems to perform ranking, as discussed in [22, 23, 104, 63, 111]. Bernstein et al. [5], Friedman et al. [28], Jia et al. [47], Lin et al. [62], Vats et al. [102], Huang et al. [44] have also considered how LLM-based recommenders can advance societal values specifically. Four capabilities of LLMs are especially salient to that task.

- **Functional language- and image-level understanding of content**. Irrespective of debates about 'true' semantic understanding [69], contemporary LLMs reliably classify content in accordance with rich evaluative criteria [112, 47, 54, 98, 45].

- **Better understanding of users** LLMs' natural language competence also enables a new kind and degree of interaction with the user, allowing models to more deeply understand users' preferences [61], as well as to provide explanations for any inferences made [30], something that traditional recommenders struggle with.

- **Reasoning and planning**. Although current models struggle in some distribution-shift scenarios [101, 60], they readily perform the core task of matching a content description to a preference profile and assigning a relevance score [5, 47, 48]. More complex pipelines embed models within agent frameworks that schedule multi-step plans [19, 31, 74, 21].

- **Tool use**. Via function-calling outputs, an LLM can invoke external software, evaluate the response, and incorporate the result into its context window for further processing [81, 43, 77]. More generally, this means that LLMs can operate as part of more complex systems, where the LLM is the executive center drawing on different software tools to enable it to achieve the tasks it has been set. Because these more complex systems involve LLMs undertaking complex sequences of action without supervision, based on its own reasoning capability, they are appropriately described as LLM agents [51].

These capacities support two broad system architectures:

**LM-assisted recommenders**. The LLM acts as a service that classifies items and scores user–item matches inside an otherwise conventional pipeline [5, 11].

**LM-agent recommenders**. The LLM holds executive control, deciding when to call auxiliary tools, retrieve new items, or update preference models (see figure above).

Regardless of architecture, an LLM-based recommender requires at least four modules. First, an inspectable, corrigible representation of each user's preferences and relevant societal values. Second, a structured record of the user's prior feed, capturing aggregate properties such as topical diversity and thematic recurrence. Third, a candidate-generation process—either platform-internal or agentic web-browsing. Fourth, a reasoning engine that combines (1)–(3) to rank items for delivery.

# 5 The Potential of an LLM-Enabled Public Sphere

The digital public sphere is dominated by firms whose business models reward large-scale behavioral surveillance and tight control over the flow of attention. If LLM-based recommenders are developed only inside those firms, they will probably inherit the same incentives [50]. Yet the capabilities of modern LLMs also make it technically possible to escape the four pathologies identified in Section 3. In this section we outline how.

## 5.1 Mitigating Mass Surveillance

LLM-based recommenders can evaluate candidate content and model user preferences directly through functional language and image understanding, and native reasoning capability, rather than through high-volume behavioral data and statistical inferences [5, 47]. Behavioral traces therefore become an avoidable externality rather than a fuel the system cannot run without. Users, or regulators, could simply forbid their collection without degrading ranking quality.

## 5.2 Diluting Concentrated Power

Because ranking quality depends mainly on the pretrained model rather than live network effects, LMAs do not automatically grow more accurate with every new user. Training frontier-scale models is still expensive, but three trends reduce the barrier to entry:

- Open-weights releases. Several near-frontier models are already available for commercial reuse.

- Distillation & synthetic data. Firms routinely shrink large checkpoints or bootstrap data from them, maintaining task performance in far smaller artifacts [14]

- Falling compute cost. Hardware and algorithmic gains [41, 17] are pushing frontier-level capability onto consumer devices; a two-year horizon for on-device operation of today's most capable models is plausible.

Together these trends will make it feasible for individuals, research labs or small companies to run personal recommenders that are not beholden to platform gatekeepers.

In addition, LLM-based *agents* can act: they detect when external tools or browser calls are useful, execute them, evaluate the result and iterate [81, 31]. A personal agent can therefore discover material scattered across blogs, pre-print servers, mailing lists or proprietary platforms—even where APIs are throttled—then aggregate it into a single feed. For example, instead of the user having to check X because 'that's where the AI papers show up,' an agent could crawl arXiv, RSS feeds, institutional repos and niche forums, returning a unified digest. Likewise, it could keep track of friends across Bluesky, Threads, LinkedIn and Mastodon, sparing users the pain of reverse-chronological doom-scrolling. Such bottom-up interoperability removes the audience lock-in on which contemporary network effects depend.[2]

On this model, platforms themselves would be likely to wither away (or else to radically change). And if everyone has their own agent advancing their own interests, then centralized control over the distribution of attention would also attenuate.

## 5.3 Beyond Narrow Behaviorism

LLMs can represent user and societal values in explicit natural-language form, accommodate higher-order preferences, and hold reflective dialog with the user [78]. Training an LLM-based agent recommender begins to resemble onboarding a human assistant: direct instructions lead to immediate generalization rather than months of implicit 'training by scrolling.' This also opens space for value-aligned nudging and for incorporating collective-risk constraints [5] instead of simply maximizing revealed preferences.

---

[2] In the EU the Digital Marketplace Act has an interoperability requirement, at present just for messaging apps. It has been agreed to discuss the feasibility of extending the requirement to social media too in future revisions. In the US the proposed ACCESS Act would also require interoperability between platforms.

## 5.4 Scaffolding Agency

Because the user model and inference chain can be surfaced in ordinary language, recommendations become explainable and corrigible. Users can ask: Why did you show me this?; edit the underlying preference statement; or override a specific suggestion. When well designed, the agent's chain-of-thought (or equivalent structured rationale) is legible even if, at present, not perfectly faithful [99]. The net effect is augmented, not diminished, attentional agency.

LM-driven recommenders can mitigate surveillance, dilute platform power, defeat lock-in, transcend narrow behaviorism and enhance user agency. The next section analyses the practical, economic and normative hurdles that must be cleared to realise this potential.

# 6 Alternative Views: Challenges for LLM Recommenders

Large-language-model recommenders promise to reduce the harms associated with today's engagement-optimized systems, but they introduce a distinct set of practical and ethical questions. We cannot resolve these issues a priori; the only decisive test is to prototype and evaluate full systems. Below we catalog the principal objections and outline why none is obviously fatal.

**Efficiency**. Even highly optimized LLMs on specialized hardware remain less efficient than recommender stacks refined over two decades.

Both algorithmic and hardware trends [41] point to rapid gains. Inference costs are already falling and are likely to approach those of traditional models within the product life-cycle of a new platform.

**Candidate Generation Requires Centralization**. Efficient first-stage retrieval appears to favor a single clearing-house that indexes content for everyone, recreating centralized power.

Retrieval can be hierarchical: lightweight local or edge models produce a coarse shortlist; stronger rerankers refine it. Falling compute costs and open-protocol infrastructure (e.g. the fediverse) mitigate the need for a single dominant index. Even if a single index were needed for candidate generation, this index need not replicate all the features of existing platforms, since agentic recommenders could absorb the other functions platforms currently perform. In general, novel infrastructure to support agents in general is needed, and can be designed around privacy and decentralization from the outset [50, 13].

**User Burden**. Systems that scaffold agency may impose unacceptable cognitive overhead; users prefer frictionless feeds.

There is no easy solution to this problem; we simply need to design UIs for LLM-based recommenders that judiciously balance user input with frictionless consumption.

**Value Representation**. LLMs have not yet shown stable, longitudinal modeling of individual preferences and values [89].

Ramos et al. [79] shows that systems that build natural-language preference profiles achieve performance comparable with matrix-factorization approaches; Zhou et al. [113] shows they can learn concise natural language preference profiles from conversation. Further progress may come from combining scaffolding and external tools to support structured generation and hierarchical representations of user values. For example, some preferences might be near-absolute constraints, some much softer. LLMs have made significant progress along this path, and promise more through the capabilities unlocked by scaling inference time compute [35, 73]. The more accurate these user models become, the more important it would be to ensure they are tightly controlled by users themselves, given the consequent privacy risk.

**Incumbent Resistance**. Platforms will adopt partial LLM solutions that preserve surveillance and power concentration, lobbying against open agents [50].

Given social media's impact on the information and communication ecosystem, regulators clearly have motive and authority to ensure genuine competition in the allocation of attention, and to support LLM recommenders [15]. And more generally, with so many companies aiming to transform personal computing entirely through LLM agents [85], the rise of LLM recommenders might ultimately be a foregone conclusion. If LLM agents become the new operating system, and the universal interface for our digital lives, recommendation will likely just fold into that broader system. We will shift from a paradigm where digital platforms are the key intermediaries controlling our digital lives, to one where LLM agents play that role [57].

These *universal intermediaries* could allow power to shift from platforms to agents—but they could also be another vector for more fine-grained control of users' lives by tech companies [50, 59].

**Security and Manipulation**: The security and safety of LLM agents raise ethical concerns. No current LLM system is immune to prompt injection [109]; hijacked agents could act on a user's behalf, armed with a detailed personal model. Trusted agents may become potent vectors for manipulation, and decentralization complicates policing [12, 13].

The threat is serious but not clearly worse than today's status quo. Robust security research, formal methods, capability bounding, and policy interventions are already under way and must be prioritized. Scaling inference compute also seems to promise enhanced adversarial robustness [108].

**Filter Bubbles**: LLM-based recommenders might create an acute version of the classic 'filter bubble' [92, 76].

This concern is likely to remain exaggerated. People can already choose to stay within their own narrow epistemic bubble, but most do not [1]. Agents can be steered toward diversity goals using the same natural-language constraints used for preference elicitation [5, 64]. Pathological designs can be forbidden by policy or protocol.

**Reduced Sociability**: Proxy agents might erode the public, interactive dimension of social media, turning it into a 'ghost town' of autonomous scrapers. Social media makes users' tastes public, and supports social taste formation. We do not consume in isolation. Proxy agents would make this more challenging.

This is a design question, not an inevitability. We can specify when to delegate exploration to agents and when to surface direct human engagement. Nothing suggests these questions are intractable.

**Upstream Ethical Costs**: Some might argue against using LLMs for *anything*. Building with LLMs entails violating others' copyright, ruining the environment, inescapably advancing the power of big tech, and so on (see e.g. [2] and the ensuing literature).

These critiques target training practices, not downstream use. Responsible data governance, green training pipelines, and diversified model supply can address the upstream harms. We reject the view that LLMs are so morally compromised as to preclude beneficial applications.

# 7 Conclusion

Early hopes that an open, networked web would democratize communication ultimately gave way to the concentrated power of a handful of platforms [3]. The emerging paradigm of LM-mediated personal computing could repeat that history; technological affordances alone cannot dissolve the structural inequities that shape today's public sphere.

Yet new capabilities expand the design space. An architecture that enables richer local inference without pervasive surveillance, that moves beyond narrow behaviorist signals, and that treats users as active principals rather than passive optimization targets represents substantive progress over the status quo. The research agenda sketched here does not eliminate every risk, but it offers a credible path toward a healthier attention ecosystem—one worth pursuing despite the challenges ahead.